\documentclass[submission, Phys]{SciPost}

\hypersetup{
    colorlinks,
    linkcolor={red!50!black},
    citecolor={blue!50!black},
    urlcolor={blue!80!black}
}

\usepackage[bitstream-charter]{mathdesign}
\DeclareSymbolFont{usualmathcal}{OMS}{cmsy}{m}{n}
\DeclareSymbolFontAlphabet{\mathcal}{usualmathcal}

\begin{document}

\begin{center}{\Large \textbf{
Topological quantum phase transition of nickelocene on Cu(100)\\
}}\end{center}

% TODO: write the author list here. Use first name (+ other initials) + surname format.
% Separate subsequent authors by a comma, omit comma and use "and" for the last author.
% Mark the corresponding author with a superscript star.
\begin{center}
G. G. Blesio\textsuperscript{1},
R. \v{Z}itko\textsuperscript{1,2},
L. O. Manuel\textsuperscript{3$\star$} and 
A. A. Aligia\textsuperscript{4}
\end{center}

\begin{center}
{\bf 1} Jo\v{z}ef Stefan Institute, Jamova 39, SI-1000 Ljubljana, Slovenia
\\
{\bf 2} Faculty of Mathematics and Physics, University of Ljubljana, Jadranska 19,
SI-1000 Ljubljana, Slovenia
\\
{\bf 3} Instituto de F\'{\i}sica Rosario (CONICET) and Facultad de Ciencias Exactas, Ingenier\'{i}a y Agrimensura, 
Universidad Nacional de Rosario, 2000 Rosario, Argentina
\\
{\bf 4} Instituto de Nanociencia y Nanotecnolog\'{\i}a CNEA-CONICET,
Centro At\'{o}mico Bariloche and Instituto Balseiro, 8400 Bariloche, Argentina
\\
${}^\star$ {\small \sf manuel@ifir-conicet.gov.ar}
\end{center}

\begin{center}
\today
\end{center}

% For convenience during refereeing (optional),
% you can turn on line numbers by uncommenting the next line:
% \linenumbers

\section*{Abstract}
{\bf
Local quantum phase transitions driven by Kondo correlations have been theoretically proposed in several magnetic nanosystems; 
however, clear experimental signatures are scant. Modeling a nickelocene molecule on a Cu(100) substrate as a two-orbital Anderson 
impurity with single-ion anisotropy coupled to two conduction bands, we find that recent scanning tunneling spectra reveal the existence 
of a topological quantum phase transition from the usual local Fermi liquid with high zero-bias conductance to 
a \textit{non-Landau} Fermi liquid, characterized by a non-trivial quantized Luttinger 
integral, with a small conductance. The effects of intermediate valence, finite 
temperature, and structural relaxation of the molecule position allow us to explain the different observed behaviors.
}

 \vspace{10pt}
 \noindent\rule{\textwidth}{1pt}
 \tableofcontents\thispagestyle{fancy}
 \noindent\rule{\textwidth}{1pt}
 \vspace{10pt}

 \section{Introduction}
 \label{sec:intro}

Quantum phase transitions (QPT) where the Kondo effect is destroyed by
competing interactions have been theoretically found in several
magnetic nanosystems~\cite{vojta06}. They were also invoked to explain
the unconventional quantum criticality in some heavy fermion 
compounds~\cite{si01}. However, experimental realizations of such
QPT~\cite{bork11} are elusive, as they require fine control of the
parameter that triggers the transition.  Due to their variety and
tunability, magnetic single molecules in contact with metal electrodes,
probed and manipulated with a scanning tunneling microscope (STM), are
distinguished candidate systems. They are being extensively
studied because of their novel properties and also their potential use
in new spintronic devices~\cite{naturefocus,cuevas,evers20,trans}.
In particular, transistors are the most important component of an integrated circuit, as they act as a switch by changing some parameter \cite{trans}. We have shown that the
the anisotropic two-channel spin-1  Kondo model
has a topological QPT changing either the spin-1 anisotropy or the Kondo exchange
parameters~\cite{blesio19,zitko21} in which the differential conductance
$\mathrm{d}I/\mathrm{d}V$ jumps between zero and the maximum possible value,
constituting an ideal switch. In the intermediate-valence regime, the
jump is still present although smaller in magnitude \cite{zitko21,blesio18}.
In this work we present theoretical evidence that shows that the system of a
nickelocene molecule on a Cu(100) substrate is the first known experimental realization
of such a molecular switch.

The double-decker nickelocene (Nc) molecules on the Cu(100) substrates
have been experimentally studied in several
articles~\cite{bache,orma1,ormaza17,orma3,verlhac,mohr}.  As the STM tip is
approached to the molecule, the tunneling regime is followed by the
contact regime~\cite{ormaza17}. While in the latter the differential
conductance $\mathrm{d}I/\mathrm{d}V$ for small bias voltage $V$ displays the
characteristic Kondo peak due to screening of the molecular magnetic
moment by conduction electrons~\cite{liang02,parks07,roch08,parks10},
in the tunneling regime $\mathrm{d}I/\mathrm{d}V$ shows a low-bias dip and finite-$V$
rise characteristic of inelastic spin-flips due to single-ion
anisotropy for a spin $S>1/2$~\cite{zitko08,zitko10}. The different
behaviors have been tentatively ascribed to a crossover in the spin of
the molecule from 1/2 in the contact regime to 1 in the tunneling
regime based on first-principle calculations~\cite{ormaza17,mohr}
that, however, neglect dynamical correlations and therefore do not
properly treat the Kondo effect. 
Moreover, the electronic structure does not change much between the
two regimes and, as admitted by the authors, the change in the
molecular charge is actually insufficient to account for the large
change in the magnetization. This calls for an alternative
interpretation without such discrepancy.

\section{Topological quantum phase transition driven by the single-ion anisotropy}
 \label{sec:tqpt}

The minimal Hamiltonian that captures the many-body physics of this
system is an impurity Anderson model in which the dominant
configuration has two holes in the Ni 3d shell occupying the nearly
degenerate orbitals with $xz$ and $yz$ symmetry, and forming a spin
$S=1$ due to the atomic Hund coupling. These triplet states are split
by the single-ion easy plane anisotropy $D$. Both localized orbitals
are equally hybridized with conduction electrons with the same
symmetry~\cite{mohr}. A recent numerical renormalization group (NRG)
study of this class of Hamiltonians~\cite{blesio18,blesio19}
demonstrated that with the increasing ratio $D/T^0_{K}$, where
$T^0_{K}$ is the $D=0$ Kondo temperature, the system undergoes a
topological quantum phase transition (TQPT) from a phase with high
impurity spectral density $\rho (\omega )$ at zero frequency $\omega $
(proportional to $\mathrm{d}I/\mathrm{d}V$ for $V=0$) to a topologically distinct phase
with a low $\rho (0)$. 
Similarly, if the hybridization between localized and conduction 
electrons $\tilde{V}$ in the Anderson model or the exchange 
coupling $J$ on the corresponding Kondo model is increased, 
$T^0_{K}$ increases, $D/T^0_{K}$ decreases and the topological 
transition from low to high $\rho (0)$ is induced.

\begin{figure}[!t]
\begin{center}
\includegraphics*[width=0.7\columnwidth]{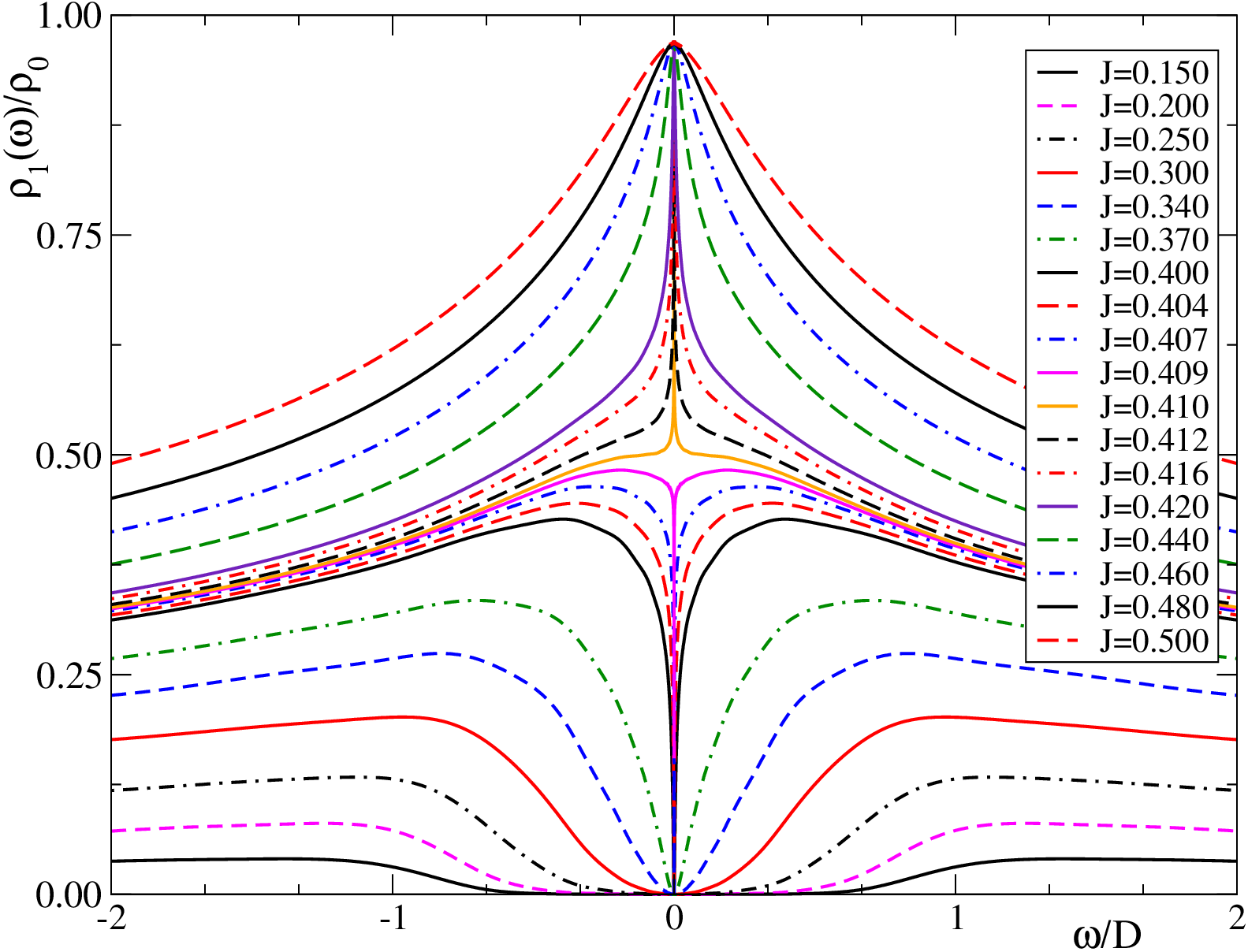}
\end{center}
\caption{Normalized spectral density of the A2CS1KM as a function of frequency 
for $D=0.042W$ and several values of the Kondo coupling $J$, given in units of the half band with $W$ of the conduction channels.}
\label{fig1}
\end{figure}

In Fig. \ref{fig1} we show the evolution of the spectral density
for the anisotropic two-channel spin-1  Kondo model (A2CS1KM)~\cite{blesio19} 
(the integer valence limit of the above Anderson model) for fixed $D$ as the Kondo exchange $J$ 
is increased. For large $J$, $\rho (\omega )$ has the usual shape expected for a Kondo resonance. 
As $J$ get closer to the critical value $J_{c}\sim 0.4095$ from above, the shape evolves to a very 
narrow peak mounted on a much broader one. For $J=J_{c}$ the narrow peak transforms suddenly into a
sharp dip, which broadens with decreasing $J$. For small $J$, $\rho(\omega)$ tends to have two well 
defined steps at $\omega=\pm D$.
A similar behavior has been
found in other strongly correlated impurity models~\cite{leo,curtin18,nishi,zitko21}.
The phase with large $D/T^0_{K}$ is a Fermi liquid characterized by a non-zero Luttinger
integral $I_{L}=\pi /2$ for each orbital and spin~\cite{blesio18,zitko21}. Since for a non-interacting 
system $I_{L}=0$, this phase breaks Landau adiabatic hypothesis. Thus for $D/T^0_K > (D/T_K^0)_c$ the system is a 
\textit{non-Landau} Fermi liquid (NLFL). Recently, we have shown~\cite{zitko21}
that this concept can explain in a unified and consistent fashion several
experiments in iron phthalocyanine (FePc) on
Au(111)~\cite{mina,hira,yang} and we have predicted the possibility of
driving that system directly through the TQPT, although such
measurement remains to be performed.

The contrasting $\mathrm{d}I/\mathrm{d}V$ spectra reported in Ref.~\cite{ormaza17}
for Nc on Cu(100) in tunneling and contact regimes coincide with those
of the two distinct phases of the model, with the tunneling regime corresponding to 
the NLFL phase. Moreover, assuming a half band with $ W =$ 1 eV, the critical anisotropy 
$D_{c}\sim 3.8$ meV~\cite{blesio18} is of the order of the reported anisotropy $D=4.2$ meV 
of the system. Then, it is
natural to assume that Nc on Cu(100) is a physical realization of the TQPT.
This would be the first
experimental realization of a TQPT to a NLFL phase. Previously, we have
identified FePc on Au(111), with a clear dip in $\mathrm{d}I/\mathrm{d}V$ at 
$V=0,$  as a NLFL~\cite{zitko21}, using a similar model as the present one with two 
non-equivalent channels~\cite{note}. Nevertheless, an abrupt jump to the conventional Fermi liquid phase
has not been reported yet in this system~\cite{note2}. 

\section{Two different behaviors of Nc/Cu(100)}
 \label{sec:two}

A recent detailed study of Nc/Cu(100) consisting of a careful
tip-molecule distance variation \cite{mohr} revealed more complex
behavior that does not quite correspond to that observed in
Ref.~\cite{ormaza17} and shown in Fig.~\ref{fig1}. Two different
types of evolution of $\mathrm{d}I/\mathrm{d}V$, probably depending on the exact
adsorption geometry, were observed. In case A, the evolution from
tunneling to contact regime is discontinuous and hysteretic, with a
jump between low and high $\mathrm{d}I/\mathrm{d}V$, seemingly skipping the regime where
a narrow peak/dip for $J\sim J_{c}$ is expected (as in
Fig.~\ref{fig1}).
In case B, observed experimentally in 1/3 of the cases, the evolution
is instead continuous, but it also lacks the narrow peak/dip
structures.

To explain these observations, we calculate the phase diagram of the
Anderson impurity model [see Eq.~(\ref{ha})] and trace the TQPT as a
function of hybridization and energy level. We find that the behavior
observed in case A can be explained by a first-order transition in the
position of the molecule that avoids parameter values near the TQPT
for which the system is less stable or unstable. For this purpose we
extend the model by including the effects of the structural relaxation
of the molecule, observed in similar systems~\cite{hira,karan} and
expected to play a role also in Nc/Cu(100). For a soft
\textquotedblleft spring\textquotedblright\ related to the molecular
displacement, we find that the system is unstable near the quantum
critical point (QCP) and a first-order TQPT takes place in which the
critical point is avoided. The continuous transition experimentally
observed in case B is due to two effects: i) the magnitude of the jump
in the spectral density at zero frequency $\rho (0)$ decreases as the
degree of intermediate valence increases and, more importantly, ii)
finite temperatures blur the narrow peaks and dips in $\rho (\omega )$
near the TQPT.

\section{Topological quantum phase transition at intermediate valence}
 \label{sec:model}

The Hamiltonian that describes the system can be written 
in the form \cite{mohr,blesio19}

\begin{eqnarray}
H  & = & \sum_{k\tau \sigma }\varepsilon _{k}c_{k\tau \sigma }^{\dagger
}c_{k\tau \sigma }+ 
\sum_{\tau \sigma }\epsilon d_{\tau \sigma }^{\dagger }d_{\tau \sigma
}+\sum_{\tau }U n_{\tau \uparrow }n_{\tau \downarrow } + \notag \\
&&+U' n_{xz}n_{yz}-J_{H}{\vec{S}}_{xz}\cdot {\vec{S}}_{yz}+DS_{z}^{2}+  
\sum_{k\tau \sigma }\left( \tilde{V} {c}_{k\tau \sigma }^{\dagger }{d}_{\tau
\sigma }+\mathrm{H.c.}\right) ,
\label{ha}
\end{eqnarray}

where $d_{\tau \sigma }^{\dagger }$ ($c^\dagger_{k \tau\sigma}$) creates a hole with energy 
$\epsilon$ ($\varepsilon_k$) in the Ni $d$ orbital $\tau$ (Cu band
$\tau$ with momentum $k$), with 
$\tau = xz, yz$. $n_{\tau \sigma }=d_{\tau \sigma}^{\dagger }d_{\tau \sigma }$ and 
$n_{\tau }=\sum_{\sigma }n_{\tau \sigma }$. The actual localized orbitals might extend in the Nc 
molecule beyond the Ni orbitals, but for simplicity we refer to them as the Ni $d$ orbitals. 
We also assume degenerate conduction bands, hybridizations $\tilde{V}$ independent of energy and channel,
and the same energy $\epsilon$ for  both orbitals~\cite{notaSOC}. 
For small hybridization and when the two-particle configuration dominates,
the model (\ref{ha}) reduces to the A2CS1KM~\cite{blesio19}.

The origin of the anisotropy $DS_{z}^{2}$ is the spin-orbit coupling term 
$H_{\text{SOC}}=\lambda \Sigma \mathbf{l}_{i}\cdot \mathbf{s}_{i}$. To
provide a qualitative understanding of the effect of $H_{\text{SOC}}$. let
us consider the effect of the $z$ components only 
$H_{\text{SOC}}^{z}=\lambda \Sigma l_{i}^{z}\cdot s_{i}^{z}$ 
on the triplet ground state 
of the configuration with two holes. This triplet, using the notation
$|SS_z \rangle$ for total spin and projection can be written as

\begin{eqnarray}
|11\rangle  &=&d_{1\uparrow }^{\dagger }d_{-1\uparrow }^{\dagger }|0\rangle ,
\quad |1-1\rangle =d_{1\downarrow }^{\dagger }d_{-1\downarrow }^{\dagger
}|0\rangle ,  \nonumber \\
|10\rangle  &=&\frac{1}{\sqrt{2}}\left( d_{1\uparrow }^{\dagger
}d_{-1\downarrow }^{\dagger }+d_{1\downarrow }^{\dagger }d_{-1\uparrow
}^{\dagger }\right) |0\rangle ,  \label{triplet}
\end{eqnarray}
where the operators with orbital projection $l_{z}=\pm 1$ are $d_{\pm
1\sigma }^{\dagger }=(\mp d_{xz\sigma }^{\dagger }-id_{yz\sigma }^{\dagger
})/\sqrt{2}$. Then
\begin{equation}
H_{\text{SOC}}^{z}|1\pm 1\rangle =0,\text{ }H_{\text{SOC}}^{z}|10\rangle
=\lambda |00\rangle ,  \label{zsoc}
\end{equation}
where the singlet $|00\rangle $ has the same form as $|10\rangle $ but with
a minus sign instead of plus in Eq. (\ref{triplet}),
and it is higher in energy by $J_H$.
Furthermore, $H^z_\text{SOC} |00\rangle = \lambda |10 \rangle$.
Thus, to second order in perturbation theory in $H_{\text{SOC}}^{z}$, 
the energy of the states $|1\pm 1\rangle $ remains the same, but the energy of $|10\rangle $ is
lowered by $D\simeq \lambda ^{2}/J_{H}$. Taking $\lambda \sim 80$ meV and 
$J_H=1.5$ eV, $D \sim 4$ meV in good agreement with the experimental 
value 4.2 meV~\cite{mohr}.

The calculations were done with the NRG Ljubljana~\cite{zitko09,nrglj} 
implementation of the numerical renormalization group method \cite{bulla,wilson}. 
The Hamiltonian was implemented with conserved total charge as well as the conservation 
of the $z$-component of the total spin, i.e., $U(1) \times U(1)$ symmetry. We kept 
up to 10000 multiplets (or up to cutoff 10 in energy units) in the truncation with the 
discretization parameter $\Lambda=4$ and averaging over $N_z=4$ different discretization 
meshes. The spectral functions were computed using the complete Fock space algorithm \cite{peters06}, 
and the resolution for the Anderson model was improved using the ``self-energy trick'' \cite{bullaself}. 

For the discussion of topological properties, we remind the reader that the
impurity spectral function {\it per} orbital and spin, at the Fermi level $(\omega
=0)$ at zero temperature, is related to the quasiparticle scattering
phase shift $\delta _{\tau \sigma }$ by~\cite{taylor,friedel,lutti,langer,lang,shiba,yoshi} 
 
\begin{equation}
\rho _{\tau \sigma }(0)=-\frac{1}{\pi }\mathrm{Im}G_{\tau \sigma }^{d}(0)=
\frac{1}{\pi \Gamma }\sin ^{2}\delta _{\tau \sigma },  \label{rho}
\end{equation}

where $G_{\tau \sigma }^{d}(\omega )=\left\langle \left\langle d_{\tau \sigma };d_{\tau \sigma }^{\dagger}\right\rangle \right\rangle $ 
is the impurity Green's function 
and $\Gamma =\pi \sum_{k}|\tilde{V}|^{2}\delta (\omega -\varepsilon _{k})$ 
is the hybridization strength, assumed independent of energy. 
In turn, the phase shift is related to the number of displaced electrons by the impurity for 
each channel and spin which, in the case of wide flat conduction density of states, coincides 
with the expectation value of the Ni occupancy for the corresponding spin and orbital. 
Therefore, in this limit, the generalized Friedel sum rule reads~\cite{blesio18,zitko21} 

\begin{equation}
\delta _{\tau \sigma }=\pi \left\langle n_{\tau \sigma }\right\rangle
+I_{\tau \sigma },\text{ }I_{\tau \sigma }=\mathrm{Im}\int_{-\infty
}^{0}d\omega \,G_{\tau \sigma }^{d}(\omega )\frac{\partial \Sigma _{\tau
\sigma }^{d}(\omega )}{\partial \omega },  \label{del}
\end{equation} 
where $\Sigma _{\tau \sigma }^{d}(\omega )$ is the impurity self energy.
In our case, by symmetry, the four integrals are equal, 
$I_{\tau \sigma }=I_{L}$. Until recently, $I_{L}=0$ was generally
assumed as a hallmark of a Fermi liquid, but several local Fermi liquids were found in 
which $I_{L}=\pm \pi /2$~\cite{blesio18,blesio19,curtin18,nishi,zitko21}.

Since $I_L=(\pi/2)\theta(D-D_c)$ \cite{blesio18,blesio19,zitko21},
Eqs.~\eqref{rho} and \eqref{del} imply that for a total Ni occupancy 
of $\left\langle n\right\rangle =4\left\langle n_{\tau \sigma }\right\rangle =2$, 
the spectral densities $\rho _{\tau \sigma }(0)$ jump from $\rho
_{0}=1/\pi \Gamma$ for $D<D_{c}$ to 0 for $D>D_{c}$.
For fixed $D$ the TQPT can be obtained by decreasing $\Gamma$ or
the exchange constant $J$, as displayed
in Fig. \ref{fig1}. 
For other occupancies, the jump in the spectral densities
at the TQPT is reduced as 
$\Delta \rho _{\tau \sigma }(0)=-\frac{1}{\pi \Gamma}\cos ( \pi \left\langle n\right\rangle /2)$.

\begin{figure}[!t]
\begin{center}
\includegraphics*[width=0.6\columnwidth]{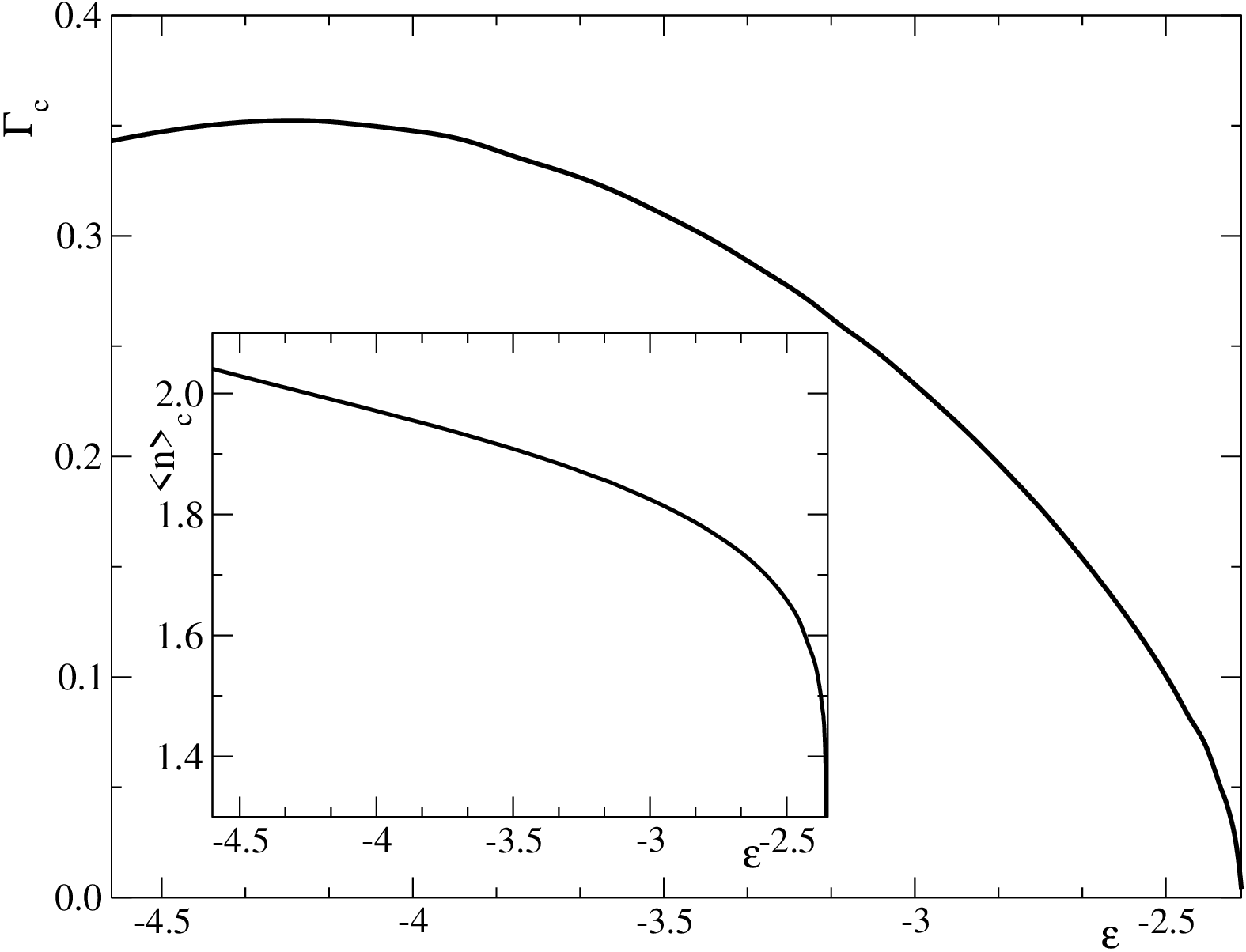}
\end{center}
\caption{Phase diagram of the Anderson model in the $\epsilon, \Gamma$ plane for 
$U=3.5$, $U'=2.5$ and $J_H=0.5$. Below (above) the curve the phase is a NLFL (usual Fermi liquid). 
Inset: occupancy of the Ni at the TQPT.}
\label{diag}
\end{figure}

Clearly this relation predicts that the jump disappears as the Ni
occupancy is reduced to one particle. More fundamentally, the
formation of the $S=1$ state requires a strong contribution of the
two-particle configuration in the ground-state wave function. This
naturally leads to the question of the very existence of the TQPT away
from half-filling. To elucidate this, we have calculated the position
of the TQPT in the $\epsilon ,\Gamma $ plane.  We take similar
parameters for the model as those reported previously~\cite{mohr} with
the half band width $W = 1$ eV as the unit of energy and a flat
density of states for the conduction electrons.  The result is
displayed in Fig.~\ref{diag}. For $\epsilon =-(U+U')/2= -4.25$,
$\left\langle n\right\rangle =2$ and $\Gamma_c$ reaches its maximum
value $\sim 350$ meV. As $\epsilon $ increases one expects that the
Kondo temperature increases for constant $\Gamma $, and then to keep
the same ratio $D/T_{K}\sim 2.6$ at the TQPT~\cite{blesio18},
$\Gamma_c $ should decrease. This decrease is more abrupt near 
$\epsilon \sim -2.5$ until for $\epsilon = \epsilon_0= -2.375$,
$\Gamma_c \rightarrow 0$ at the TQPT. For larger $\epsilon $, the
system is always in the topologically trivial ($I_L=0$) phase. 

Note that the end point of the TQPT $(\epsilon ,\Gamma )=(\epsilon _{0},0)$ is the point of degeneracy of the ground state configurations of the
isolated molecule ($\Gamma =0$) between the two-particle state with $S_{z}=0$
[$|10 \rangle$ given by Eq. (\ref{triplet}) with energy $2\epsilon +U^{\prime }-J_{H}/4$] and the four one-particle
states ($d_{\pm 1 \sigma }^{\dagger }|0\rangle $  
with energy $\epsilon $). A TQPT at this point might be expected on
general physical grounds. If the one-particle configuration is that of
lowest energy, the splitting of the orbitals is physically relevant and
cannot be neglected for a realistic description (as for FePc \cite{note}).
Neglecting the excited doublet, 
the model mixes the doublet of the states 
$d_{1\uparrow }^{\dagger}|0\rangle $,  
and $d_{-1\downarrow }^{\dagger }|0\rangle $ with the triplet
given by Eqs. (\ref{triplet}).
For $D=0$, the model has been solved exactly \cite{bethe} 
and the ground state
is a triplet. For arbitrary $D$ and small $\Gamma $,  a Schrieffer-Wolf
transformation like that performed in Ref.~\cite{bethe} leads to the
following Kondo interaction 
\begin{equation}
H_{I}=\left( \frac{-2 \tilde{V}^{2}}{\Delta E+D}+\frac{\tilde{V}^{2}}{\Delta E}\right)
S_{d}^{z}S_{c}^{z}-\frac{\tilde{V}^{2}}{\Delta E}\left(
S_{d}^{x}S_{c}^{x}+S_{d}^{y}S_{c}^{y}\right) ,  \label{hint}
\end{equation}
where $\vec{S}_{d}$ and $\vec{S}_{c}$ are the total spin of the localized
and conduction electrons respectively, and $\Delta E$ is the excitation
energy from the one-particle GS configuration to the $S_{z}=0$ triplet
excited state. According to poor man's scaling \cite{ander}, for $D>0$, the ground state
is a singlet and corresponds to an ordinary Fermi liquid as we found. If
instead, the two-particle configuration dominates, the model is equivalent
to the A2CS1KM \cite{blesio19} with exchange $J\rightarrow 0$ which is a NLFL.

In the inset of Fig. \ref{diag} we show the Ni occupancy $\left\langle n\right\rangle$
along the TQPT line. It lies always above $\sim 1.5$ . For $\left\langle n\right\rangle =1.5$, the jump in the spectral
density is reduced by a factor 0.71. 
Therefore, while the TQPT is present with some degree of intermediate
valence, it seems that the two-particle triplet Ni configuration should be the dominant 
one for the TQPT to be observed.

We now discuss the evolution of $\mathrm{d}I/\mathrm{d}V$ for type B~\cite{mohr}, where
a jump the spectral density at the TQPT is not observed, but there is
a rather continuous evolution as the STM tip is pressed against the
surface, increasing $\Gamma $. As discussed above, the jump can be
reduced up to 30\% by the effects of intermediate valence. Such
interpretation is supported by the measured conductances that saturate
well below the unitary limit (see Fig. 4 in Ref.~\cite{mohr}).
Nevertheless, an even more important effect is that due to the finite
temperature of the experiment ($T=$ 4.5 K): the narrow dip/peak
near the TQPT (see Fig.~\ref{fig1}) is blurred by temperature and,
furthermore, at finite temperature $\mathrm{d}I/\mathrm{d}V$ is not proportional to
$\rho _{\tau\sigma }(\omega )$ but to its convolution with the
derivative of the Fermi function (see for example Ref.~\cite{zitko21})
that further broadens sharp features near the TQPT. 

\begin{figure}[!t]
\begin{center}
\includegraphics*[width=\columnwidth]{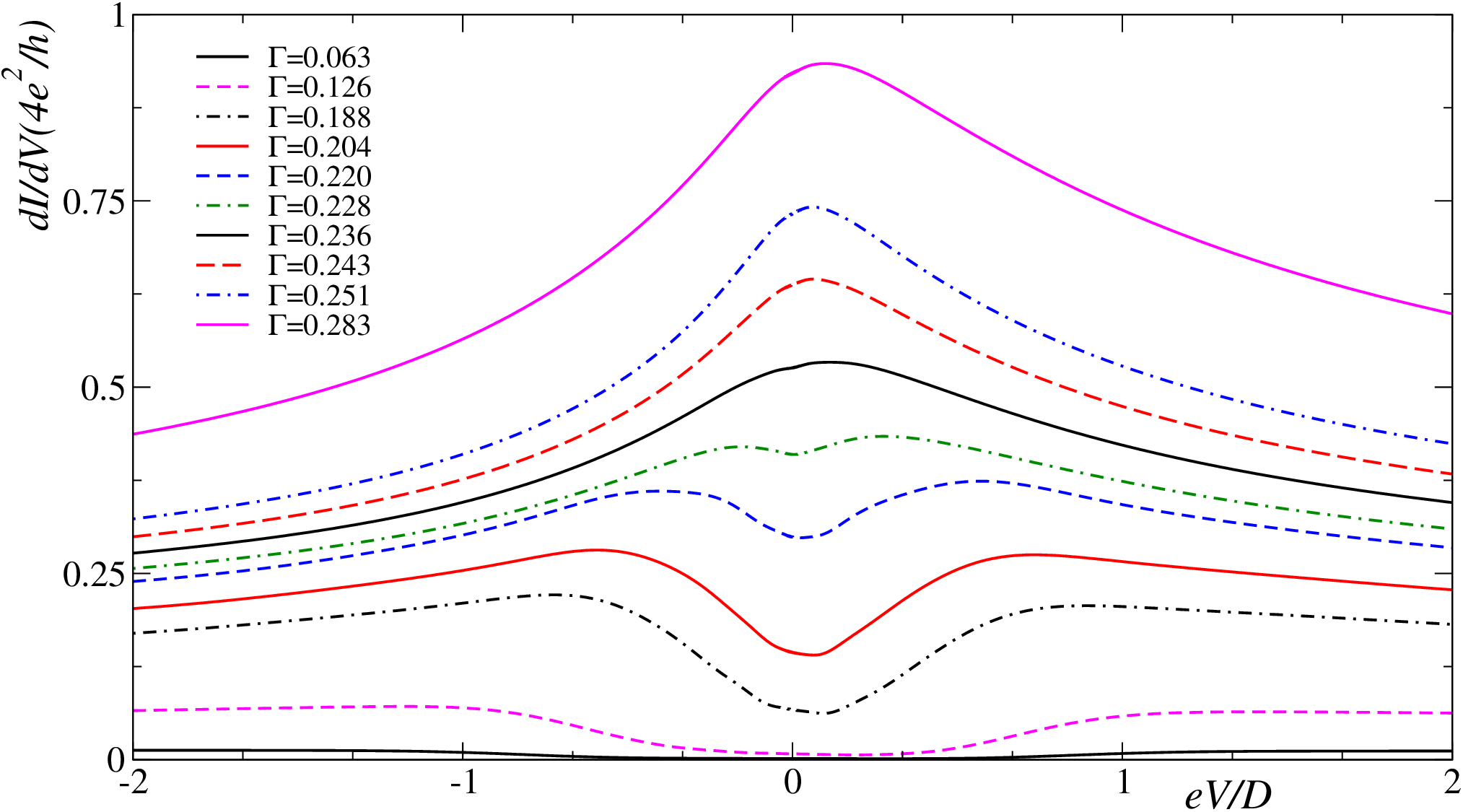}
\end{center}
\caption{Differential conductance as a function of voltage for different 
values of $\Gamma$. Other parameters are $U=3.5$, $U'=2.5,$ $J_H=0.5$,
$\epsilon=-3.0$, $T=0.0005$.}
\label{figb}
\end{figure}

In Fig. \ref{figb} we show the evolution of the differential conductance as 
$\Gamma $ is increased at a finite temperature. 
For small $\Gamma $, $\mathrm{d}I/\mathrm{d}V$ has a dip mounted on a broader peak. As $\Gamma $ increases, 
the dip narrows, but in contrast to
the case of zero temperature, the minimum of the dip increases and a very
sharp dip like that of Fig. \ref{fig1} is absent. 
The effect of temperature on the dip is similar to that shown 
in Fig. 3 of Ref.~\cite{zitko21}.
For larger $\Gamma$ the dip disappears and the magnitude of 
$\mathrm{d}I/\mathrm{d}V$ near zero voltage increases. 
A sharp peak like that of Fig. 1 for $J$ slightly above the 
transition value $J_c$ is also absent, mainly as an effect 
of finite temperature. 
Therefore, the main features of Fig. 5 (b) of Ref.~\cite{mohr}
are reproduced. 
Our curves seem somewhat more asymmetric than the experimental ones. 
This can be corrected by decreasing $\epsilon$.

\section{Structure relaxation of the nickelocene molecule and first-order transition}
 \label{sec:relaxation}

To explain the evolution of $\mathrm{d}I/\mathrm{d}V$ for the experimental situation of
type A (see Section 3 for the explanation of types A and B), 
we need to extend the model to permit the
structural relaxation of the molecule. Clearly $\Gamma $ increases as
the STM tip approaches to the Cu(100) surface, and (as in other
systems~\cite{hira,karan}) the Nc molecule accommodates itself to the
tip position, mainly tilting its axis relative to the surface. This
shift in the position of the molecule, which we denote as $\eta $,
also affects $\Gamma$. On general physical grounds one expects, to
leading order in $\eta$,
\begin{equation}
\Gamma =\Gamma _{0}+a\eta ,  \label{deltaeta}
\end{equation}
where $\Gamma _{0}$, which depends on the position of the STM\ tip, is the
value of $\Gamma $ for $\eta =0$ and $a$ is a constant. The on-site energy 
$\epsilon $ might depend slightly on $\eta $ but this does not affect the 
main argument below. The total Hamiltonian $H_{t}$ should also include the
elastic energy due to the displacement $\eta$,
\begin{equation}
H_{t}=H+\frac{1}{2}K\eta ^{2},  \label{ht}
\end{equation}
where $K>0$ is another constant. The optimum value of $\eta $ for each 
$\Gamma _{0}$ is obtained by minimizing the total energy $E_{t}=\left\langle
H_{t}\right\rangle =E(\Gamma )+b\eta ^{2}$, where $E=\left\langle
H\right\rangle $. Differentiating this equation one has
\begin{eqnarray}
\frac{dE_{t}}{d\eta } =\frac{\partial E}{\partial \Gamma }a+K\eta, ~~ 
\frac{d^{2}E_{t}}{d\eta ^{2}} =\frac{\partial ^{2}E}{\partial \Gamma ^{2}}a^{2}+K.  
\label{deri}
\end{eqnarray}
In order to have a locally stable minimum at the value $\eta =\eta _{1}$ for
which $dE_{t}/d\eta =0$, one needs that  $ d^{2}E_{t}/d\eta ^{2} > 0$ for $\eta
=\eta _{1}$. If, however, for some value $\Gamma =\Gamma _{u}$, \ $\partial
^{2}E/\partial \Gamma ^{2}<-K/a^{2}$, the position $\eta =\eta _{1}$ of the
molecule is unstable and, as $\Gamma _{0}$ is varied, the system has a first-order 
transition between two states, one with $\Gamma <$ $\Gamma _{u}$ and another one 
with $\Gamma >$ $\Gamma _{u}$, avoiding the value $\Gamma =\Gamma _{u}$. 

\begin{figure}[!t]
\begin{center}
\includegraphics*[width=0.8\columnwidth]{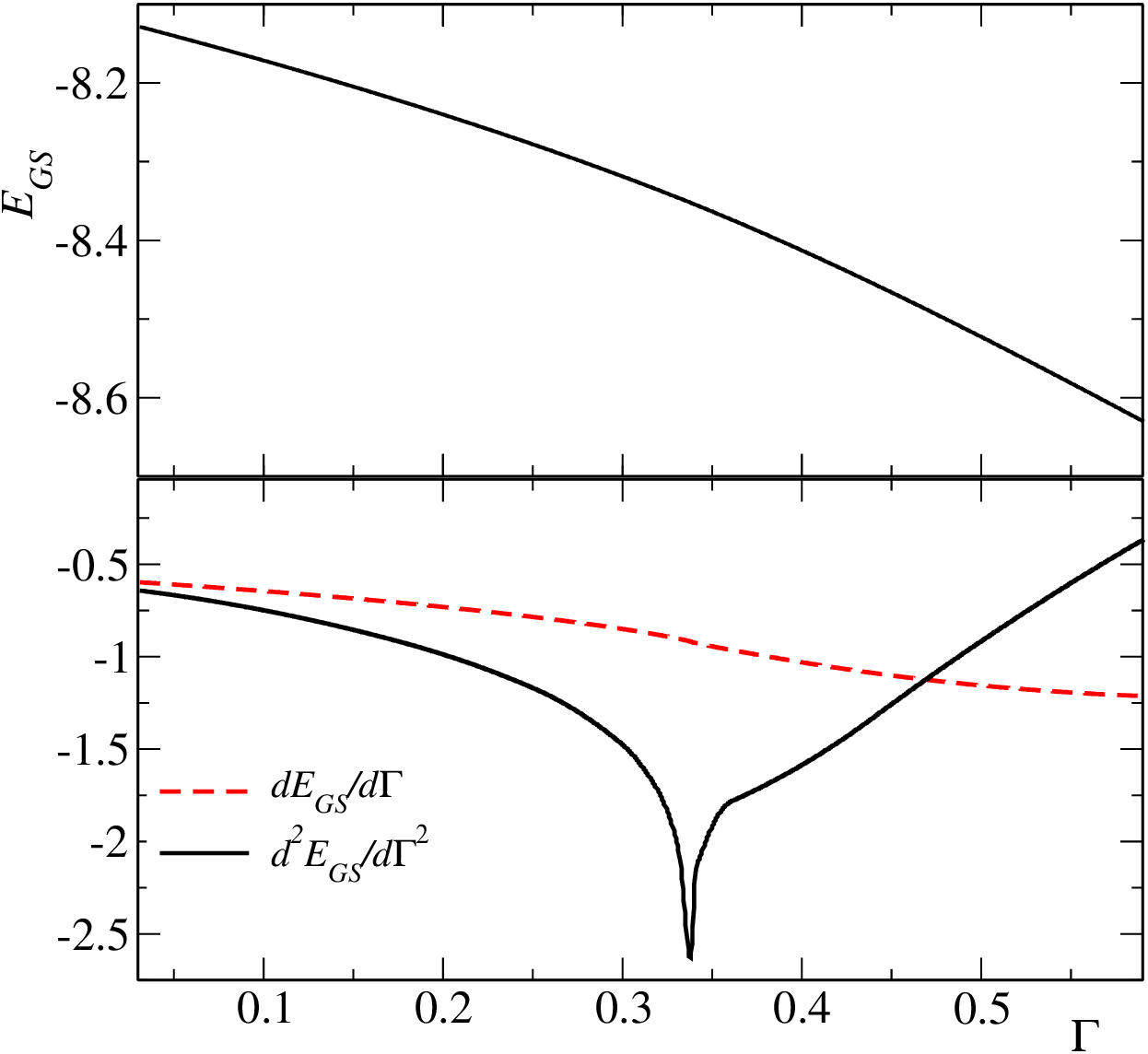}
\end{center}
\caption{Ground-state energy and its first and second derivatives 
as a function of $\Gamma$ for $U=3.5$, $U'=2.5$ and $J_H=0.5$, and $\epsilon=-3.8$.}
\label{figene}
\end{figure}

To analyze this possibility in our system, we have calculated the
ground state energy $E(\Gamma )$ and numerically differentiated it,
see Fig.~\ref{figene}. Remarkably, $\partial
^{2}E/\partial \Gamma ^{2} $ is negative and large in magnitude at 
the value $\Gamma=\Gamma_c$ that corresponds to the TQPT.
Therefore, one expects that for a soft enough spring constant
($K<a^{2}|\partial ^{2}E/\partial \Gamma ^{2}|$), as the STM tip is
pressed into the molecule, there is a first-order transition
from the NLFL phase to the usual Fermi liquid phase, avoiding the values
$\Gamma \sim \Gamma _{c}$. This agrees with the experimental
observations reported in Ref.~\cite{ormaza17} and Fig. 5 (a) of
Ref.~\cite{mohr} for systems of type A. For the cases of type B, it is
likely that relaxation of the molecule is inhibited by a larger spring
constant $K$. In the contact regime, the Kondo resonance is
much broader for the type B situation than for the type A, suggesting
an upright molecular configuration in case B and a tilted one in case
A~\cite{mohr}.

 \section{Conclusion}

We have uncovered a topological quantum phase transition
(TQPT) in published tunneling spectra of nickelocene molecules on
Cu(100).  The TQPT is controlled by the distance between the STM tip
and the surface.  We have modeled the system as a spin-1 two-channel
Anderson model with easy-plane magnetic anisotropy, which has a TQPT
between a non-trivial topological \textit{non-Landau} Fermi liquid
phase in the tunneling regime to a topologically trivial Fermi liquid
phase in the contact regime. The transition extends to the
zero-hybridization limit at the degeneracy point between the
two-particle configuration and the one-/three-particle configuration.
We find that the different behaviors of the differential conductance
$\mathrm{d}I/\mathrm{d}V$ as a function of voltage can be explained in terms of this
TQPT.  The transition is characterized by a jump in $\mathrm{d}I/\mathrm{d}V$ at $V=0$
at zero temperature. However, in most experimental observations (case
A) there is a hysteretic jump in $\mathrm{d}I/\mathrm{d}V$ also at finite voltage which
can be explained by the structural relaxation of the molecule.  In the
absence of this relaxation and at finite temperature, the change in
$\mathrm{d}I/\mathrm{d}V$ is continuous as the tip of the STM is approached to the
molecule (case B). To our knowledge, this is the first experimental realization of the 
TQPT and might have applications in molecular electronics. We hope that our work will 
stimulate further research and help the interpretation of ongoing experiments in 
similar systems.

\section*{Acknowledgements}
R\v{Z} acknowledges the support of the Slovenian Research Agency (ARRS)
under P1-0044 and P1-0416. GGB is also supported by the Slovenian 
Research Agency (ARRS) under P1-0044 and J1-2458.
GGB and LOM are supported by PIP No. 3220 of CONICET, Argentina. AAA
is supported by PICT 2017-2726 and PICT 2018-01546 of the ANPCyT, Argentina.

% TODO: include author contributions
% \paragraph{Author contributions}
% This is optional. If desired, contributions should be succinctly described in a single short paragraph, using author initials.

% TODO: include funding information
% \paragraph{Funding information}

\nolinenumbers


\begin{thebibliography}{99}

\bibitem{vojta06} 
M. Vojta, 
{\it Impurity quantum phase transitions}, 
Philos. Mag. \textbf{86}, 1807 (2006), 
\doi{10.1080/14786430500070396}.

\bibitem{si01} 
Q. Si, S. Rabello, K. Ingersent and J. Lleweilun Smith, 
{\it Locally critical quantum phase transitions in strongly correlated metals}, 
Nature \textbf{413}, 804 (2001), 
\doi{10.1038/35101507}. 

\bibitem{bork11} 
J. Bork, Y. Zhang, L. Diekhöner, L. Borda, P. Simon, J. Kroha, P. Wahl and K. Kern, 
{\it A tunable two-impurity Kondo system in an atomic point contact}, 
Nature Phys. \textbf{7}, 901 (2011), 
\doi{10.1038/nphys2076}. 

\bibitem{naturefocus} 
S. V. Aradhya and L. Venkataraman,
{\it  Single-molecule junctions beyond electronic transport}, 
Nature Nanotechnology \textbf{8} 399 (2013),
\doi{10.1038/nnano.2013.91}. 

\bibitem{cuevas} 
J. C. Cuevas and E. Scheer, 
\textit{Molecular Electronics: An Introduction to Theory and Experiment}, 
World Scientific, Singapore (2010), 
\doi{10.1142/10598}.

\bibitem{evers20} 
F. Evers, R. Koryt\'ar, S. Tewari and J. van Ruitenbeek,
{\it Advances and challenges in single-molecule electron transport}, 
Rev. Mod. Phys. \textbf{92}, 035001 (2020),
\doi{10.1103/RevModPhys.92.035001}. 

\bibitem{trans} 
P. Th. Mathew and F. Fang, 
{\it Advances in Molecular Electronics: A Brief Review},
Engineering \textbf{4}, 760 (2018), 
\doi{10.1016/j.eng.2018.11.001}. 

\bibitem{blesio19} 
G. G. Blesio, L. O. Manuel, P. Roura-Bas and A. A. Aligia, 
{\it Fully compensated Kondo effect for a two-channel spin $S = 1$ impurity}, 
Phys. Rev. B \textbf{100}, 075434 (2019), 
\doi{10.1103/PhysRevB.100.075434}. 

\bibitem{zitko21} 
R. \v{Z}itko, G. G. Blesio, L. O. Manuel and A. A. Aligia,
{\it Iron phthalocyanine on Au(111) is a ``non-Landau'' Fermi liquid},
Nature Commun. \textbf{12}, 6027 (2021), 
\doi{10.1038/s41467-021-26339-z}. 

\bibitem{blesio18} 
G. G. Blesio, L. O. Manuel, P. Roura-Bas and A. A. Aligia, 
{\it Topological quantum phase transition between Fermi liquid phases in an Anderson impurity model}, 
Phys. Rev. B \textbf{98}, 195435 (2018), 
\doi{10.1103/PhysRevB.98.195435}. 

\bibitem{bache} 
N. Bachellier, M. Ormaza, M. Faraggi, B. Verlhac, M. V\'{e}rot, T. Le Bahers, M.-L. Bocquet and L. Limot,
{\it  Unveiling nickelocene bonding to a noble metal surface}, 
Phys. Rev. B \textbf{93}, 195403 (2016), 
\doi{10.1103/PhysRevB.93.195403}. 

\bibitem{orma1} 
M. Ormaza, R. Robles, N. Bachellier, P. Abufager, N. Lorente and L. Limot,
{\it  On-surface engineering of a magnetic organometallic nanowire}, 
Nano Lett. \textbf{16}, 588 (2016), 
\doi{10.1021/acs.nanolett.5b04280}. 

\bibitem{ormaza17} 
M. Ormaza, P. Abufager, B. Verlhac, N. Bachellier, M.-L. Bocquet, N. Lorente and L. Limot, 
{\it Controlled spin switching in a metallocene molecular junction}, 
Nat. Commun. \textbf{8}, 1974 (2017), 
\doi{10.1038/s41467-017-02151-6}. 

\bibitem{orma3} 
M. Ormaza, N. Bachellier, M. N. Faraggi, B. Verlhac, P. Abufager, P. Ohresser, L. Joly, M. Romeo, F. Scheurer, M.-L. Bocquet, N. Lorente and L. Limot, 
{\it  Efficient spin-flip excitation of a nickelocene molecule}, 
Nano Lett. \textbf{17}, 1877 (2017), 
\doi{10.1021/acs.nanolett.6b05204}. 

\bibitem{verlhac} 
B. Verlhac, N. Bachellier, L. Garnier, M. Ormaza, P. Abufager, R. Robles, M.-L. Bocquet, M. Ternes, N. Lorente and L. Limot,
{\it Atomic-scale spin sensing with a single molecule at the apex of a scanning tunneling microscope},
Science \textbf{366}, 623 (2019), 
\doi{10.1126/science.aax8222}. 

\bibitem{mohr} 
M. Mohr, M. Gruber, A. Weismann, D. Jacob, P. Abufager, N. Lorente and R. Berndt, 
{\it Spin dependent transmission of nickelocene-Cu contacts probed with shot noise}, 
Phys. Rev. B \textbf{101}, 075414 (2020), 
\doi{10.1103/PhysRevB.101.075414}. 

\bibitem{liang02} 
W. Liang, M. P. Shores, M. Bockrath, J. R. Long and H. Park,
{\it  Kondo resonance in a single-molecule transistor}, 
Nature \textbf{417}, 725 (2002), 
\doi{10.1038/nature00790}. 

\bibitem{parks07} 
J. J. Parks, A. R. Champagne, G. R. Hutchison, S. Flores-Torres, H. D. Abru\~{n}a and D. C. Ralph, 
{\it Tuning the Kondo Effect with a Mechanically Controllable Break Junction}, 
Phys. Rev. Lett. \textbf{99}, 026601 (2007), 
\doi{10.1103/PhysRevLett.99.026601}. 

\bibitem{roch08} 
N. Roch, S. Florens, V. Bouchiat, W. Wernsdorfer and F. Balestro, 
{\it Quantum phase transition in a single-molecule quantum dot}, 
Nature \textbf{453}, 633 (2008), 
\doi{10.1038/nature06930}. 

\bibitem{parks10} 
J. J. Parks, A. R. Champagne, T. A. Costi, W. W. Shum, A. N. Pasupathy, E. Neuscamman, S. Flores-Torres, P. S. Cornaglia, 
A. A. Aligia, C. A. Balseiro, G. K.-L. Chan, H. D. Abru\~{n}a and D. C. Ralph,
{\it Mechanical Control of Spin States in Spin-1 Molecules and the Underscreened Kondo Effect}, 
Science \textbf{328}, 1370 (2010), 
\doi{10.1126/science.1186874}. 

\bibitem{zitko08} 
R. \v{Z}itko, R. Peters and T. Pruschke, 
{\it Properties of anisotropic magnetic impurities on surfaces}, 
Phys. Rev. B \textbf{78}, 224404 (2008), 
\doi{10.1103/PhysRevB.78.224404}. 

\bibitem{zitko10} 
R. \v{Z}itko and Th. Pruschke, 
{\it Many-particle effects in adsorbed magnetic atomswith easy-axis anisotropy: the case of Fe on theCuN/Cu(100) surface}, 
New J. Phys. \textbf{12}, 063040 (2010), 
\doi{10.1088/1367-2630/12/6/063040}. 

\bibitem{leo} 
L. De Leo and M. Fabrizio, 
{\it Spectral properties of a two-orbital Anderson impurity model across a non-Fermi-liquid fixed point},
Phys. Rev. B \textbf{69}, 245114 (2004), 
\doi{10.1103/PhysRevB.69.245114}. 

\bibitem{curtin18} 
O. J. Curtin, Y. Nishikawa, A. C. Hewson and D. J. G. Crow,
{\it  Fermi liquids and the Luttinger theorem}, 
J. Phys. Commun. \textbf{2}, 031001 (2018), 
\doi{10.1088/2399-6528/aab00e}. 

\bibitem{nishi} 
Y. Nishikawa, O. J. Curtin, A. C. Hewson and D. J. G. Crow,
{\it Magnetic field induced quantum criticality and the Luttinger sum rule}, 
Phys. Rev. B \textbf{98}, 104419 (2018), 
\doi{10.1103/PhysRevB.98.104419}. 

\bibitem{mina} 
E. Minamitani, N. Tsukahara, D. Matsunaka, Y. Kim, N. Takagi and M. Kawai,
{\it  Symmetry-Driven Novel Kondo Effect in a Molecule}, 
Phys. Rev. Lett. \textbf{109}, 086602 (2012), 
\doi{10.1103/PhysRevLett.109.086602}. 

\bibitem{hira} 
R. Hiraoka, E. Minamitani, R. Arafune, N. Tsukahara, S. Watanabe, M. Kawai and N. Takagi, 
{\it Single-molecule quantum dot as a Kondo simulator}, 
Nature Commun. \textbf{8}, 16012 (2017), 
\doi{10.1038/ncomms16012}. 

\bibitem{yang} 
K. Yang, H. Chen, Th. Pope, Y. Hu, L. Liu, D. Wang, L. Tao, W. Xiao, X. Fei, Y-Y. Zhang, H-G Luo, S. Du, T. Xiang, W. A. Hofer and H-J. Gao,
{\it  Tunable giant magnetoresistance in a single-molecule junction}, 
Nature Commun. \textbf{10}, 1038 (2019), 
\doi{10.1038/s41467-019-11587-x}. 

\bibitem{note} A detailed justification of the A2CS1KM
with inequivalent channels used in Ref. \cite{zitko21} is given in
A. A. Aligia, {\it Low-energy physics for an iron phthalocyanine molecule on Au(111)},
Phys. Rev. B \textbf{105}, 205114 (2022), \doi{10.1103/PhysRevB.105.205114}.

\bibitem{note2}  Applying a magnetic field in FePc, the dip
is transformed gradually to a peak \cite{yang} because, as a consequence of the symmetry
breaking, the topological protection of the Luttinger integral is lost \cite{zitko21}.

\bibitem{karan} 
S. Karan, D. Jacob, M. Karolak, Ch. Hamann, Y. Wang, A. Weismann, A. I. Lichtenstein and R. Berndt, 
{\it Shifting the Voltage Drop in Electron Transport Through a Single Molecule}, 
Phys. Rev. Lett. \textbf{115}, 016802 (2015), 
\doi{10.1103/PhysRevLett.115.016802}. 

\bibitem{notaSOC} Actually a splitting is induced by a tilt of the molecule
and by spin-orbit coupling (SOC) \cite{note}. However we have verified
that a splitting of 80 meV (of the order of the SOC for Ni), 
only reduces by near 7\% the value of $\tilde{V}$ at the TQPT. 

\bibitem{zitko09} 
R. \v{Z}itko and T. Pruschke, 
{\it Energy resolution and discretization artifacts in the numerical renormalization group}, 
Phys. Rev. B \textbf{79}, 085106 (2009), 
\doi{10.1103/PhysRevB.79.085106}. 

\bibitem{nrglj} NRG Ljubljana, \url{https://github.com/rokzitko/nrgljubljana} and \url{http://nrgljubljana.ijs.si/}.

\bibitem{bulla} 
R. Bulla, T. Costi and T. Pruschke, 
{\it The numerical renormalization group method for quantum impurity systems}, 
Rev. Mod. Phys. \textbf{80}, 395 (2008), 
\doi{10.1103/RevModPhys.80.395}. 

\bibitem{wilson} 
K. G. Wilson, 
{\it The renormalization group: Critical phenomena and the  Kondo problem}, 
Rev. Mod. Phys. \textbf{47}, 773 (1975), 
\doi{10.1103/RevModPhys.47.773}. 

\bibitem{peters06} 
R. Peters, Th. Pruschke and F. B. Anders, 
{\it A numerical renormalization group approach to Green's functions for quantum impurity models}, 
Phys. Rev. B \textbf{74}, 245114 (2006), 
\doi{10.1103/PhysRevB.74.245114}. 

\bibitem{bullaself} 
R. Bulla, A. C. Hewson and Th. Pruschke, 
{\it Numerical renormalization group calculations for the self-energy of the impurity Anderson model}, 
J. Phys.: Condens. Matter \textbf{10}, 8365 (1998), 
\doi{10.1088/0953-8984/10/37/021}.

\bibitem{taylor} 
J. R. Taylor, 
\textit{Scattering Theory}, 
John Wiley \& Sons, New York (1972). ISBN: 0486450139.


\bibitem{friedel} 
J. Friedel, 
{\it XIV. The distribution of electrons around impurities in monovalent metals}, 
Philos. Mag. \textbf{43}, 153 (1952), 
\doi{10.1080/14786440208561086}. 

\bibitem{lutti} 
J. M. Luttinger, 
{\it Fermi Surface and Some Simple Equilibrium Properties of a System of Interacting Fermions}, 
Phys. Rev. \textbf{119}, 1153 (1960), 
\doi{10.1103/PhysRev.119.1153}. 

\bibitem{langer} 
J. S. Langer and V. Ambegaokar, 
{\it Friedel Sum Rule for a System of Interacting Electrons}, 
Phys. Rev. \textbf{121}, 1090 (1961), 
\doi{10.1103/PhysRev.121.1090}. 

\bibitem{lang} 
D. C. Langreth, 
{\it Friedel Sum Rule for Anderson's Model of Localized Impurity States}, 
Phys. Rev. \textbf{150}, 516 (1966), 
\doi{10.1103/PhysRev.150.516}. 

\bibitem{shiba} 
H. Shiba, 
{\it The Korringa Relation for the Impurity Nuclear Spin-Lattice Relaxation in Dilute Kondo Alloys}, 
Progress of Theoretical Physics \textbf{54}, 967 (1975), 
\doi{10.1143/PTP.54.967}. 

\bibitem{yoshi} 
A Yoshimori and A Zawadowski, 
{\it Restricted Friedel sum rules and Korringa relations as consequences of conservation laws}, 
J. Phys. C \textbf{15}, 5241 (1982), 
\doi{10.1088/0022-3719/15/25/015}.

\bibitem{bethe} 
A. A. Aligia, C. A. Balseiro and C. R. Proetto,
{\it Integrability of a general model for intermediate valence},
Phys. Rev. B \textbf{33}, 6476 (1986), 
\doi{10.1103/PhysRevB.33.6476}. 

\bibitem{ander} 
P. W. Anderson, 
{\it A poor man's derivation of scaling laws for the Kondo problem},
J. Phys. \textbf{C3}, 2439 (1970), 
\doi{10.1088/0022-3719/3/12/008}. 


\end{thebibliography}
\end{document}